\documentclass[twocolumn,showpacs,preprintnumbers,amsmath,amssymb,pra,nofootinbib]{revtex4-1}

\usepackage[dvipdfmx]{graphicx}
\usepackage{dcolumn}
\usepackage{bm}
\usepackage{color}

\begin{document}

\title{Quantum Computation Based on Quantum Adiabatic Bifurcations of Kerr-Nonlinear Parametric Oscillators}

\author{Hayato Goto}
\affiliation{Frontier Research Laboratory, 
Corporate Research \& Development Center, 
Toshiba Corporation, Kawasaki, Kanagawa 212-8582, Japan}

\begin{abstract}
\noindent Quantum computers with Kerr-nonlinear parametric oscillators (KPOs) have recently been proposed by the author and others.
Quantum computation using KPOs is based on quantum adiabatic bifurcations of the KPOs,
which lead to quantum superpositions of coherent states, such as Schr\"{o}dinger cat states.
Therefore, these quantum computers are referred to as ``quantum bifurcation machines (QbMs)."
QbMs can be used for qauntum adiabatic optimization and universal quantum computation.
Superconducting circuits with Josephson junctions, Josephson parametric oscillators (JPOs) in particular, 
are promising for physical implementation of KPOs.
Thus, KPOs and QbMs offer not only a new path toward the realization of quantum bits (qubits) and quantum computers,
but also a new application of JPOs.
Here we theoretically explain the physics of KPOs and QbMs, 
comparing them with their dissipative counterparts.
Their physical implementations with superconducting circuits are also presented.
\end{abstract}

\maketitle

\section{Introduction}

Stimulated by commercial quantum computers developed by D-Wave systems~\cite{Harris2008a,Harris2010a,Johnson2011a,Lanting2014a,Boixo2014a,Denchev2016a}, 
which are based on quantum annealing~\cite{Kadowaki1998a,Brooke1999a,Santoro2002a,Das2008a} 
or adiabatic quantum computation~\cite{Farhi2000a,Farhi2001a,Kaminsky2004a,Amin2008a,Albash2018a},
hardware devices for combinatorial optimization
have attracted much attention.
D-Wave's quantum annealers are Ising machines, that is,
designed for finding ground states of the Ising model~\cite{Barahona1982a}.
Since many combinatorial optimization problems can be mapped to the Ising problem~\cite{Lucas2014a}, 
fast Ising machines are expected to be useful for various real-world problems, including 
very-large-scale integrated (VLSI) circuit design~\cite{Barahona1988a}, 
computational biology~\cite{Perdomo2008a,Perdomo-Ortiz2012a,Li2018a},
classification problems~\cite{Neven2008a,Neven2009a},
scheduling or planning problems~\cite{Rieffel2015a,Venturelli2015a},
drug design~\cite{Sakaguchi2016a}, 
financial portfolio management~\cite{Rosenberg2016a},
and traffic-flow optimization~\cite{Neukart2017a}.
Other kinds of Ising machines have also been proposed and developed using 
laser pulses~\cite{Utsunomiya2011a,Takata2012a,Taketa2014a,Utsunomiya2015a,Wang2013a,Marandi2014a,
Haribara2015a,Haribara2016a,Takata2016a,Takesue2016a,Inagaki2016a,McMahon2016a,Haribara2017a,
Takata2015a,Shoji2017a,Yamamura2017a,Yamamoto2017a},
electromechanical resonators~\cite{Mahboob2016a},
CMOS circuits~\cite{Yamaoka2016a},
and magnetic devices~\cite{Mizushima2017a}.

The first kind of laser-based Ising machine
was proposed by Utsunomiya et al.~\cite{Utsunomiya2011a} and later developed by them~\cite{Takata2012a,Taketa2014a,Utsunomiya2015a}.
This machine uses an injection-locked laser network and 
the ``up" and ``down" states of each Ising spin 
are represented by the polarizations~\cite{Utsunomiya2011a,Takata2012a,Taketa2014a} or 
phases~\cite{Utsunomiya2015a} of each mode.
Importantly, 
Utsunomiya et al. introduced a new operational principle 
referred to as the \textit{minimum-gain principle}~\cite{Utsunomiya2011a,Wang2013a}.
The Ising energy (the cost function of the problem) 
is mapped to the total loss in the laser network.
Then, the mode configuration with the minimum total loss requires 
the minimum gain for oscillation (the lowest threshold).
Hence it oscillates most stably among all the configurations.
Thus, we will obtain the ground state of the Ising model from the mode configuration of the steady state, 
assuming that the system chooses the most stable state with the minimum total loss
as the steady state.

The second kind of laser-based Ising machine, 
which is called a \textit{coherent Ising machine} (CIM), 
was proposed by Wang et al.~\cite{Wang2013a}
and experimentally realized by them~\cite{Marandi2014a}.
This is also based on the minimum-gain principle,
but uses a network of optical parametric oscillators (OPOs).
An OPO is suitable for Ising machines,
because it has two stable oscillating states above the threshold,
which naturally represent an Ising spin.
In other words,
an OPO is an optical implementation of the ``parametron."~\cite{Goto1959a}
Large-scale CIMs have already been realized~\cite{Inagaki2016a,McMahon2016a} 
using a measurement-feedback technique~\cite{Haribara2015a,Haribara2016a}.

Interestingly,
operation of the CIM was explained in an opposite way to annealing approaches~\cite{Marandi2014a}.
That is, in contrast to annealing, 
the CIM finds the ground state of the Ising model by increasing the gain gradually
and finding the mode configuration with the lowest threshold,
where the direction of the search is ``upward" in the energy landscape~\cite{Marandi2014a}.

The proposal of CIMs suggested parametric-oscillator implementations of Ising machines.
Mahboob et al.~\cite{Mahboob2016a} proposed an Ising machine with electromechanical parametric resonators
as a phononic counterpart of optical CIMs.

Since the above minimum-gain principle inevitably requires losses,
which usually degrade quantum superpositions and lead to decoherence,
it seems difficult for machines based on this principle to realize quantum computation in the standard sense.
To make the minimum-gain principle compatible with quantum coherence,
Goto theoretically proposed an Ising machine composed of Kerr-nonlinear parametric oscillators (KPOs)~\cite{Goto2016a}.
A KPO is a parametric oscillator with large Kerr (or Duffing) nonlinearity and without losses 
in the ideal case.
Here, Kerr (Duffing) nonlinearity provides a quadratic (quartic) energy term
with respect to oscillation power (amplitude).
While in an OPO, the threshold is determined by one-photon loss and the oscillation is stabilized by two-photon loss, 
in a KPO, a detuning and the Kerr effect play the roles of the threshold and the stabilization, respectively.
Since the detuning and the Kerr effect are described by the corresponding Hamiltonian terms,
a KPO operates without dissipation and can maintain quantum coherence.
That is, a KPO is a nondissipative counterpart of an OPO.

A KPO can deterministically generate Schr\"{o}dinger cat states~\cite{Leonhardt,Haroche2013a,Wineland2013a,Ourjoumtsev2006a,Ourjoumtsev2007a,
Deleglise2008a,Vlastakis2013a,Leghtas2015a,Wang2016a,Sychev2017a} 
(quantum superpositions of coherent states) 
via quantum adiabatic evolution increasing the pumping rate gradually.
Here, coherent states are eigenstates of annihilation operators~\cite{Leonhardt}
and regarded as the ``most classical" states of light.
The cat-state generation is understood as a result of a \textit{quantum adiabatic bifurcation} 
of the KPO~\cite{Goto2016a}.

By introducing nondissipative linear couplings between KPOs according to the coupling coefficients
of a given Ising model,
the Ising energy is approximately mapped to the threshold of the KPO network.
Thus, the minimum-gain principle is generalized to nondissipative oscillator networks.
However, the two approaches rely on different mechanisms.
In the case of an OPO network,
the state of the system converges to an attractive steady state corresponding to 
a low Ising energy due to network losses.
On the other hand, 
the KPO network has no losses and therefore has no attractive states.
Instead, the KPO network can find the ground state of the Ising model
via quantum adiabatic evolution increasing the pumping rate gradually.
That is, convergence of the KPO network to an optimal solution
is guaranteed by the quantum adiabatic theorem.
This is called \textit{bifurcation-based adiabatic quantum computation}~\cite{Goto2016a},
to distinguish it from conventional adiabatic quantum computation or quantum annealing,
where quantum fluctuation terms are decreased gradually.
(Puri et al.~\cite{Puri2017a} reformulated adiabatic quantum computation using KPOs
in a similar manner to quantum annealing.)
Machines based on quantum adiabatic bifurcations of KPOs
are called \textit{quantum bifurcation machines} (QbMs)~\cite{Goto2018a} 
(with a lower-case ``b" to distinguish this from quantum Boltzmann machines (QBMs)~\cite{Amin2018a}).

KPOs that can generate cat states deterministically have not been experimentally realized to date.
As suggested in Ref.~\citenum{Goto2016a},
superconducting circuits with Josephson junctions are promising
for implementing KPOs,
because large Kerr effects can be realized using the nonlinearity of 
Josephson junctions~\cite{Kirchmair2013a,Rehak2014a}
and parametric modulation can be implemented easily by modulating the magnetic flux 
through a dc superconducting quantum interference device 
(SQUID)~\cite{Yamamoto2008a,Bourassa2012a,Wustmann2013a,Krantz2013a,Eichler2014a,Lin2014a,Krantz2016a}.
Superconducting devices using such parametric modulation are known as
Josephson parametric amplifiers (JPAs) or Josephson parametric oscillators (JPOs).
KPOs and QbMs offer a new application of JPOs.
Nigg et al.~\cite{Nigg2017a} and Puri et al.~\cite{Puri2017a}
proposed superconducting-circuit implementations for QbMs with \textit{all-to-all connectivity}.
The scheme proposed by Puri et al.~\cite{Puri2017a}, 
which is based on the Lechner-Hauke-Zoller (LHZ) scheme~\cite{Lechner2015a,Rocchetto2016a,Chancellor2017a}
proposed for all-to-all connected quantum annealers,
is particularly promising.
The four-body constraint required for the LHZ scheme,
which is a technical difficulty in this scheme,
is naturally realized by four-wave mixing in a Josephson junction.
Using the technique to transform the four-body constraint to a three-body one~\cite{Leib2016a},
Zhao et al. proposed an alternative architecture with three-dimensional microwave cavities~\cite{Zhao2017a}.

Single KPOs with small dissipation have been studied theoretically 
in the field of quantum nonlinear dynamics~\cite{Dykman,Marthaler2006a,Dykman2011a}.
These studies have led to ``quantum heating,"
which is a heating process among quasienergy states by dissipation,
where quasienergy states are eigenstates of the Hamiltonian in a rotating frame
and in the rotating-wave approximation.
Recently, Goto et al.~\cite{Goto2018a} generalized the concept of quantum heating
from a single nonlinear oscillator to multiple coupled nonlinear oscillators
through the study of dissipative QbMs.
This has opened new possibilities for the application of KPO networks,
such as Boltzmann sampling for Boltzmann machine learning 
in the field of artificial intelligence~\cite{Goto2018a,Amin2018a,MacKay}.

KPOs and QbMs have also opened new possibilities for standard gate-based quantum computers.
After the proposal of the Ising machine with KPOs~\cite{Goto2016a},
Goto~\cite{Goto2016b} and Puri et al.~\cite{Puri2017b}
proposed gate-based universal quantum computation using two oscillating states of a KPO as a qubit.
(Other kinds of quantum computers with cat states or similar bosonic codes have also been proposed and 
developed~\cite{Nigg2014a,Mirrahimi2014a,Albert2016a,Ofek2016a,Zhang2017a,Rosenblum2018a,Chou2018a,Puri2018a,Rosenblum2018b}.)
The fact that QbMs can perform universal quantum computation is significant, because
it suggests that classical simulation of QbMs is extremely hard for the following reason.
If QbMs are efficiently simulated using classical computers by any method,
universal quantum computation can be simulated classically through QbM simulation.
On the other hand, from quantum computational complexity theory,
it is strongly believed that even non-universal quantum computation cannot be simulated classically~\cite{Bremner2011a}.
This leads to the hardness of classical simulation of QbMs.
In contrast, there has not been such evidence for the hardness of classical simulation of CIMs so far.

In this paper,
we describe the physics of KPOs and QbMs 
in comparison with OPOs and CIMs.
Comparisons of a single KPO and a KPO network (QbM)
with a single OPO and an OPO network (CIM)
are summarized in Tables~\ref{table-summary-1} and \ref{table-summary-2}, respectively.
(The present models for an OPO and an OPO network are minimum ones
for direct comparisons with a KPO and a KPO network.
See Refs.~\citenum{Wang2013a,Marandi2014a,Takata2016a,Takesue2016a,Inagaki2016a,McMahon2016a,Haribara2015a,Haribara2016a,Haribara2017a,Takata2015a,Yamamura2017a,Shoji2017a,Yamamoto2017a} 
for more sophisticated or realistic models for OPOs and CIMs.)
From these comparisons, 
the KPO is a nondissipative (imaginary) counterpart of the OPO.
The remainder of this paper is organized as follows.
In Sect.~\ref{sec-KPO}, we describe the dynamics of a single KPO using its quantum and classical models.
In Sect.~\ref{sec-network},
we start with a summary of the theoretical aspects of a KPO network (QbM), and then 
we present simulation results for two coupled KPOs and four-spin Ising machines.
In Sect.~\ref{sec-SC}, we present superconducting-circuit implementations of a KPO and QbMs.
In Sect.~\ref{sec-universal},
we briefly explain how to realize a universal gate set for QbMs.
Finally, a summary and outlook are provided in Sect.~\ref{sec-conclusion}.

\begin{widetext}

\begin{table}
\caption{
Comparison of a single KPO and a single OPO. 
$\mathcal{H}$ and $\mathcal{L}$ are the Hamiltonian for a KPO and the Liouvillian for an OPO, respectively.
(We use the unit ${\hbar=1}$.)
$p$ is the parametric pumping rate.
$K$ and $\Delta$ are the Kerr coefficient and the detuning, respectively, for the KPO.
$\kappa$ and $\kappa_2$ are the one-photon and two-photon loss rates, respectively, for the OPO.
The classical models are defined with a complex amplitude ${\alpha = x + iy}$.
$H$ and $E$ are the Hamiltonian for the classical KPO and the energy for the classical OPO, respectively.
$p_{\mathrm{th}}$ denotes the thresholds (bifurcation points) for the classical models of the oscillators.
``Oscillation amplitudes" are amplitudes corresponding to stable fixed points in the classical models.}
\label{table-summary-1}
\begin{center}
\begin{tabular}{c|c|c}
\hline
 \multicolumn{1}{c|}{}  &  \multicolumn{1}{c|}{Single KPO} &   \multicolumn{1}{c}{Single OPO} \\
\hline
 & Schr\"{o}dinger equation  & Master equation \\
 & 
$\displaystyle |\dot{\psi} \rangle = -i \mathcal{H} |\psi \rangle$ &
$\displaystyle \dot{\rho}= \mathcal{L} \rho = -i [\mathcal{H},\rho] + \mathcal{L}_1 \rho + \mathcal{L}_2 \rho $ \\
\shortstack{Quantum \\ model} & 
$\displaystyle {\mathcal{H}=\frac{K}{2} a^{\dagger 2} a^2 + \Delta a^{\dagger} a - \frac{p}{2} \left( a^{\dagger 2} + a^2\right)}$ &
$\displaystyle {\mathcal{H}=i \frac{p}{2} \left( a^{\dagger 2} - a^2\right)}$ \\
 & & 
$\displaystyle \mathcal{L}_1 \rho = \kappa \left( 2 a \rho a^{\dagger} - a^{\dagger} a \rho - \rho a^{\dagger} a \right)$ \\
 & & 
$\displaystyle \mathcal{L}_2 \rho = \frac{\kappa_2}{2} \left( 2 a^2 \rho a^{\dagger 2} - a^{\dagger 2} a^2 \rho - \rho a^{\dagger 2} a^2 \right)$ \\
\cline{1-3}
  &  
$\displaystyle \dot{\alpha}= i \left( p \alpha^* - \Delta \alpha - K|\alpha |^2 \alpha \right)$ &
$\displaystyle \dot{\alpha}= p \alpha^* - \kappa \alpha - \kappa_2|\alpha |^2 \alpha$ \\
 \shortstack{Classical \\ model} & 
$\displaystyle \dot{x}=\frac{\partial H}{\partial y}= \left[ p+\Delta+K(x^2+y^2)\right]y$ &
$\displaystyle \dot{x}=-\frac{\partial E}{\partial x}= \left[ p-\kappa-\kappa_2(x^2+y^2)\right]x$ \\
 & 
$\displaystyle \dot{y}=-\frac{\partial H}{\partial x}= \left[ p-\Delta-K(x^2+y^2)\right]x$ &
$\displaystyle \dot{y}=-\frac{\partial E}{\partial y}= -\left[ p+\kappa+\kappa_2(x^2+y^2)\right]y$ \\
\cline{1-3}
\shortstack{ \\ Classical \\ energy} & 
$\displaystyle {H=-\frac{p}{2} \left( x^2-y^2 \right) + \frac{\Delta}{2} \left( x^2+y^2 \right) + \frac{K}{4} \left( x^2+y^2 \right)^2}$ & 
$\displaystyle {E=-\frac{p}{2} \left( x^2-y^2 \right) + \frac{\kappa}{2} \left( x^2+y^2 \right) + \frac{\kappa_2}{4} \left( x^2+y^2 \right)^2}$ \\
\cline{1-3}
 Threshold & $p_{\mathrm{th}}=\Delta$~(Detuning) & $p_{\mathrm{th}}=\kappa$~(One-photon loss) \\
\cline{1-3}
 Stabilization & Kerr effect & Two-photon loss \\
\cline{1-3}
 \shortstack{ \\ Oscillation \\ amplitudes} & 
$\displaystyle \pm \sqrt{\frac{p-\Delta}{K}}$ &
$\displaystyle \pm \sqrt{\frac{p-\kappa}{\kappa_2}}$ \\
\hline
\end{tabular}
\end{center}

\caption{
Comparison of a KPO network (QbM) and an OPO network (CIM).
$\mathcal{H}_i^{(1)}$ and $\mathcal{L}_i^{(1)}$ are the Hamiltonian for the $i$-th KPO 
and the Liouvillian for the $i$-th OPO, respectively,
defined as $\mathcal{H}$ and $\mathcal{L}$ in Table~\ref{table-summary-1}.
$\mathcal{H}_c$ and $\mathcal{L}_c$ describe the couplings for KPOs and OPOs, respectively.
$J_{i,j}$ is the coupling coefficient between the $i$-th and $j$-th Ising spins in a given Ising model.
$\xi_0$ is a positive constant with the dimension of frequency.
$H_i^{(1)}$ and $E_i^{(1)}$ are the classical Hamiltonian for the $i$-th KPO 
and the classical energy for the $i$-th OPO, respectively,
defined as $H$ and $E$ in Table~\ref{table-summary-1}.
$p_{\mathrm{th}}$ denotes the thresholds (bifurcation points) for the classical models of the oscillator networks.
$\lambda_{\mathrm{max}}$ is the maximum eigenvalue of coupling matrix $J$.}
\label{table-summary-2}
\begin{center}
\begin{tabular}{c|c|c}
\hline
 \multicolumn{1}{c|}{}  &  \multicolumn{1}{c|}{KPO network (QbM)} &   \multicolumn{1}{c}{OPO network (CIM)} \\
\hline
  & Schr\"{o}dinger equation & Master equation \\
 \shortstack{Quantum \\ model} & 
$\displaystyle |\dot{\psi} \rangle = -i \sum_{i=1}^N \mathcal{H}^{(1)}_i |\psi \rangle -i\mathcal{H}_c |\psi \rangle$ &
$\displaystyle \dot{\rho}= \sum_{i=1}^N \mathcal{L}^{(1)}_i \rho + \mathcal{L}_c \rho $ \\
 & 
$\displaystyle {\mathcal{H}_c= - \xi_0 \sum_{i=1}^N \sum_{j=1}^N J_{i,j} a^{\dagger}_i a_j}$ &
$\displaystyle \mathcal{L}_c \rho = -\xi_0 \sum_{i=1}^N \sum_{j=1}^N J_{i,j} \left( 2 a_i \rho a^{\dagger}_j - a^{\dagger}_j a_i \rho - \rho a^{\dagger}_j a_i \right)$ \\
\cline{1-3}
  &  
$\displaystyle \dot{\alpha}_i= i \left( p \alpha^*_i - \Delta \alpha_i - K|\alpha_i |^2 \alpha_i 
+\xi_0 \sum_{j=1}^N J_{i,j} \alpha_j \right)$ &
$\displaystyle \dot{\alpha}_i= p \alpha^*_i - \kappa \alpha_i - \kappa_2|\alpha_i |^2 \alpha_i +\xi_0 \sum_{j=1}^N J_{i,j} \alpha_j$ \\
\shortstack{Classical \\ model} & 
$\displaystyle \dot{x}_i=\frac{\partial H}{\partial y_i}= \left[ p+\Delta+K(x_i^2+y_i^2)\right]y_i -\xi_0 \sum_{j=1}^N J_{i,j} y_j$ &
$\displaystyle \dot{x}_i=-\frac{\partial E}{\partial x_i}= \left[ p-\kappa-\kappa_2(x_i^2+y_i^2)\right]x_i +\xi_0 \sum_{j=1}^N J_{i,j} x_j$ \\
 & 
$\displaystyle \dot{y}_i=-\frac{\partial H}{\partial x_i}= \left[ p-\Delta-K(x_i^2+y_i^2)\right]x_i +\xi_0 \sum_{j=1}^N J_{i,j} x_j$ &
$\displaystyle \dot{y}_i=-\frac{\partial E}{\partial y_i}= -\left[ p+\kappa+\kappa_2(x_i^2+y_i^2)\right]y_i +\xi_0 \sum_{j=1}^N J_{i,j} y_j$ \\
\cline{1-3}
\shortstack{ \\ \\ Classical \\ energy} & 
$\displaystyle {H= \sum_{i=1}^N H^{(1)}_i - \frac{\xi_0}{2} \sum_{i=1}^N \sum_{j=1}^N J_{i,j} (x_i x_j + y_i y_j)}$ & 
$\displaystyle {E= \sum_{i=1}^N E^{(1)}_i - \frac{\xi_0}{2} \sum_{i=1}^N \sum_{j=1}^N J_{i,j} (x_i x_j + y_i y_j)}$ \\
\cline{1-3}
 Threshold & $p_{\mathrm{th}}=\Delta- \xi_0 \lambda_{\mathrm{max}}$ 
 & $p_{\mathrm{th}}=\kappa- \xi_0 \lambda_{\mathrm{max}}$ \\
\hline
\end{tabular}
\end{center}
\end{table}

\end{widetext}

\section{Physics of a KPO}
\label{sec-KPO}

In a frame rotating at half the pump frequency, $\omega_p/2$, 
and in the rotating-wave approximation, 
a quantum model of a KPO is given 
by the Schr\"{o}dinger equation with the Hamiltonian $\mathcal{H}$ 
in Table~\ref{table-summary-1}~\cite{Goto2016a,Goto2016b},
where $a^{\dagger}$ and $a$ are the creation and annihilation operators, respectively,
for the KPO.
The Hamiltonian includes three terms corresponding to
a Kerr effect, a detuning, and parametric pumping (two-photon driving).
Here we assume a positive Kerr coefficient (${K>0}$).
If ${K<0}$ as in the case of 
Josephson parametric oscillators~\cite{Puri2017a,Nigg2017a,Puri2017b,Bourassa2012a,Wustmann2013a,Krantz2013a,Eichler2014a,Lin2014a,Krantz2016a},
we redefine the Hamiltonian $\mathcal{H}$ by flipping the overall phase,
after which we obtain the same results.
The detuning $\Delta$ is defined as ${\Delta = \omega_{\mathrm{KPO}} - \omega_p/2}$,
where $\omega_{\mathrm{KPO}}$ is the one-photon resonance frequency of the KPO.
Hereafter we assume ${\Delta>0}$.
This is a natural choice because the Kerr effect leads to larger detunings for larger amplitudes, and 
therefore stabilizes the oscillation.
The case where ${\Delta<0}$ is briefly discussed later.
This corresponds to the case in Ref.~\citenum{Puri2017a}, 
where the Kerr coefficient is negative and the detuning is positive.

A corresponding model for an OPO is given by the master equation in Table~\ref{table-summary-1},
where the Kerr effect and the detuning are replaced by two-photon and one-photon losses, respectively.
In the case of OPOs,
the two-photon loss leads to larger losses for larger amplitudes, and thus stabilizes the oscillation.
The phase of the pump amplitude is also redefined
so that the oscillation phases are the same as those for the KPO.

Here we introduce classical models~\cite{Goto2016a} corresponding to the quantum models,
which are useful for grasping the dynamics of the oscillators.
The classical models in Table~\ref{table-summary-1} are derived as follows.
Using the Schr\"{o}dinger and master equations,
the expectation values of the annihilation operator $a$ satisfy
\begin{align}
&\mathrm{KPO}:~
\langle \dot{a} \rangle
=
i \left( p \langle a^{\dagger} \rangle - \Delta \langle a \rangle
-K \langle a^{\dagger} a^2 \rangle \right),
\label{eq-expectation-KPO}
\\
&\mathrm{OPO}:~
\langle \dot{a} \rangle
=
p \langle a^{\dagger} \rangle - \kappa \langle a \rangle
-\kappa_2 \langle a^{\dagger} a^2 \rangle,
\label{eq-expectation-OPO}
\end{align}
where $\langle O \rangle$ represents the expectation value of an operator $O$ and 
the dot denotes differentiation with respect to time $t$.
These equations clearly show that 
the KPO is an imaginary counterpart of the OPO,
and the detuning and the Kerr effect 
correspond to the one-photon and two-photon losses, respectively.
When the state is near to a coherent state $|\alpha \rangle$,
moments are approximated as 
$\langle a^{\dagger m} a^n \rangle \approx \alpha^{* m} \alpha^n$,
where the asterisk denotes complex conjugation.
Thus, the classical models in Table~\ref{table-summary-1} are derived 
from Eqs.~(\ref{eq-expectation-KPO}) and (\ref{eq-expectation-OPO}).

The classical model for the KPO
can be reformulated as a classical Hamiltonian dynamical system
with the Hamiltonian $H$ in Table~\ref{table-summary-1}.
Thus, $H$ is conserved in this model,
provided that the system parameters are constant.
On the other hand,
in the classical model for the OPO,
the energy $E$ in Table~\ref{table-summary-1} 
of the same form as $H$ decreases monotonically,
because
\begin{align}
\dot{E} \approx \frac{\partial E}{\partial x} \dot{x} +\frac{\partial E}{\partial y} \dot{y} 
=-\left( \dot{x}^2 + \dot{y}^2 \right) < 0,
\end{align}
where the term proportional to $\dot{p}$ has been disregarded.
Thus, the state of the OPO varies towards a local minimum of $E$.

\begin{figure}[b]
	\includegraphics[width=6.5cm]{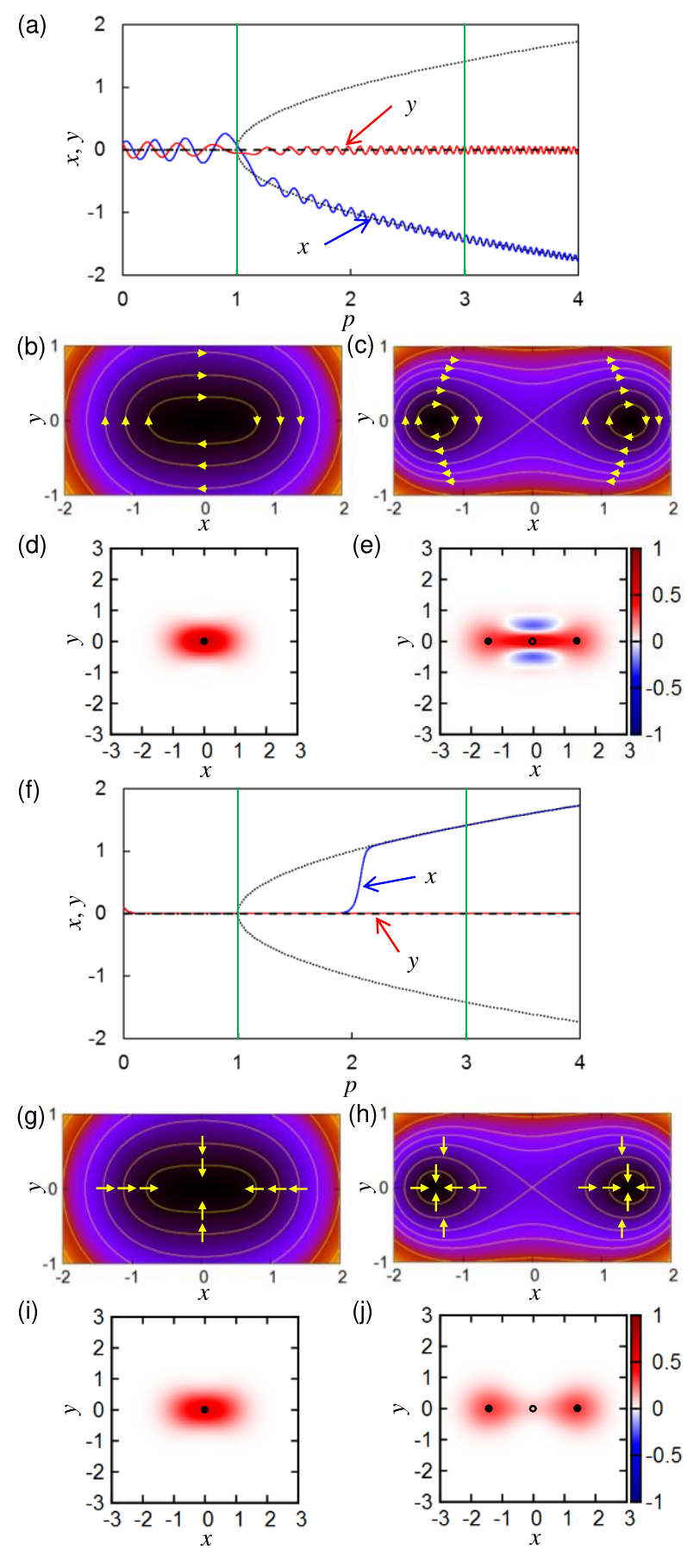}
	\caption{Simulation results for a KPO and an OPO.
	Parameters are set as ${K=\Delta=1}$ and ${\kappa_2 = \kappa =1}$
	($K$ or $\kappa_2$ is the unit of frequency), and 
	$p$ is increased linearly from 0 to 4 at ${t=100}$.
	(a) Time evolution of a classical KPO.
	(b) and (c) Phase portraits of the classical KPO at ${p=1}$ (b) and ${p=3}$ (c). 
	(d) and (e) Wigner functions of the corresponding quantum KPO at ${p=1}$ (d) and ${p=3}$ (e). 
	(f)--(j) are the corresponding results for an OPO.
	Stable and unstable fixed points in the classical models are represented, respectively, 
	by the dotted and dashed lines in (a) and (f), and 
	by the filled and open circles in the figures for the Wigner functions, 
	Vertical lines in (a) and (f) indicate the values of $p$ for the phase portraits and the Wigner functions.}
	\label{fig-single-KPO}
\end{figure}

Before explaining the quantum models,
we discuss the dynamics of the classical models
to grasp the essence of these oscillators.
Simulation results for the classical models of KPOs and OPOs are shown 
in Figs.~\ref{fig-single-KPO}(a) and \ref{fig-single-KPO}(f), respectively.
In these simulations,
we set the parameters as ${K=\Delta=1}$ and ${\kappa_2 = \kappa =1}$
($K$ or $\kappa_2$ is the unit of frequency) and 
increase the pumping rate $p$ linearly from 0 to 4 at the final time ${t=100}$.
The initial values are set as ${x=y=0.1}$.
In these figures,
the dotted and dashed lines represent stable and unstable fixed points~\cite{Strogatz}, respectively,
where the fixed points are defined by ${\dot{x}=\dot{y}=0}$.
That is, the fixed points correspond to extrema of the Hamiltonian $H$ or energy $E$.
At ${p=\Delta}$ or ${p=\kappa}$,
a single fixed point at the origin becomes two stable fixed points and an unstable fixed point,
which is called a pitchfork bifurcation~\cite{Strogatz}.
Stable fixed points are given by the ``oscillation amplitudes" in Table~\ref{table-summary-1}.
As shown in these figures,
the system follows or converges to one of the two stable fixed points.
In the case of an OPO,
this is natural because the equations of motion
lead to a monotonical decrease of $E$, and therefore
the state approaches one of the local minima of $E$, which correspond to the stable fixed points.
These dynamics are depicted in the phase portraits~\cite{Strogatz} 
in Figs.~\ref{fig-single-KPO}(g) and \ref{fig-single-KPO}(h),
where the lines represent contours of $E$.
On the other hand,
the energy of the KPO is conserved and the state varies along a contour of $H$,
as shown in Figs.~\ref{fig-single-KPO}(b) and \ref{fig-single-KPO}(c).
The dynamics of the KPO in Fig.~\ref{fig-single-KPO}(a) 
can be explained using an adiabatic invariant in classical mechanics
~\cite{Goto2016a,Landau,Goldstein,Arnold}.
The adiabatic invariant is defined as the area enclosed by the trajectory in the phase space.
When $p$ varies slowly,
the adiabatic invariant is kept at a small value,
which holds only near the local minima of the Hamiltonian.
Consequently, the system follows one of them.
We refer to this process as an \textit{adiabatic bifurcation}.

Next, we perform similar simulations using the quantum models,
where both the systems are initially in vacuum.
The Wigner functions corresponding to the phase portraits 
are shown in Fig.~\ref{fig-single-KPO},
where filled and open circles represent stable and unstable fixed points, respectively, in the classical models.
Here the Wigner function is a quasi-probability distribution 
defined as $\displaystyle {W(\alpha ) = 
\frac{2}{\pi} \mbox{Tr} \! \left[ D(-\alpha) \rho D(\alpha) P \right]}$~\cite{Leonhardt,Deleglise2008a,Goto2016a}
(${\alpha =x+iy}$),
where $\displaystyle {D(\alpha)=\exp \! \left( {\alpha a^{\dagger} - \alpha^* a} \right)}$ is the displacement operator
and 
$\displaystyle {P=\exp \! \left( i\pi a^{\dagger} a \right)}$ is the parity operator.
At the classical bifurcation points (thresholds),
the two oscillators become similar squeezed states,
as shown in Figs.~\ref{fig-single-KPO}(d) and \ref{fig-single-KPO}(i).
The large quantum fluctuations in the $x$ direction
may be useful when searching for optimal solutions in QbMs~\cite{Goto2016a} and 
CIMs~\cite{Yamamoto2017a}.

In the case of the OPO,
the quantum state above the threshold
is a mixed state of two coherent states corresponding to the classical stable fixed points,
as shown in Fig.~\ref{fig-single-KPO}(j).
This can be understood from the master equation as follows.
When ${\kappa =0}$, the master equation in Table~\ref{table-summary-1}
has exactly two steady states $|{\pm \alpha_{SS}} \rangle$,
where $|{\pm \alpha_{SS}} \rangle$ are coherent states and 
${\alpha_{SS} =\sqrt{p/\kappa_2}}$.
When there exists small one-photon loss,
coherence between the two states is lost due to the loss,
and consequently the steady state becomes a mixed state of the two coherent states.
The amplitudes become smaller due to the loss,
which correspond to the classical stable fixed points
(the ``oscillation amplitudes" in Table~\ref{table-summary-1}).
(If the one-photon loss is negligibly small,
the relaxation due to the two-photon loss leads to steady cat states~\cite{Milburn}.
This has been experimentally demonstrated using superconducting circuits~\cite{Leghtas2015a,Touzard2018a}.)

On the other hand, the Wigner function for a KPO above the threshold in Fig.~\ref{fig-single-KPO}(e)
shows an interference fringe between two coherent states,
which means that the two states are superposed.
That is, a KPO above the threshold becomes an even cat state,
${|\alpha_S \rangle + |{- \alpha_S} \rangle}$,
where ${\pm \alpha_S}$ correspond to the two classical stable fixed points
(the ``oscillation amplitudes" in Table~\ref{table-summary-1}).
Thus in quantum mechanics, the system follows two bifurcating branches ``simultaneously."~\cite{Goto2016a} 
This intriguing process is referred to as a \textit{quantum adiabatic bifurcation}~\cite{Goto2016a}.

Its mechanism is explained as follows.
For simplicity, we first assume no detuning.
Then the Hamiltonian $\mathcal{H}$ in Table~\ref{table-summary-1}
can be rewritten as
\begin{align}
\mathcal{H}
=\frac{K}{2} \left( a^{\dagger 2} - \frac{p}{K} \right) \left( a^{2} - \frac{p}{K} \right),
\end{align}
where a c-number term has been dropped.
Note that this is a positive semidefinite operator and 
that ${\mathcal{H}|{\pm \alpha_S} \rangle = 0}$.
Hence the two coherent states are exactly degenerate ground states of $\mathcal{H}$.
From the quantum adiabatic theorem,
the final state becomes a ground state of the final Hamiltonian
as long as the variation of $p$ is sufficiently slow.
Since the Hamiltonian is symmetric under the parity inversion ${a\to -a}$,
parity is conserved.
The vacuum state has even parity, and therefore
we obtain the even cat state ${|\alpha_S \rangle+ |{- \alpha_S} \rangle}$
via the quantum adiabatic evolution.
When ${\Delta>0}$,
the vacuum state is a single ground state of the initial Hamiltonian.
If the final value of $p$ is large as compared to $\Delta$,
the detuning term can be treated as a perturbation.
Assuming the final states are approximately composed of coherent states,
the amplitudes are determined by the variational method minimizing the final energy~\cite{Goto2018a}.
The resultant amplitudes are exactly the same as the stable fixed points of the classical model
(the ``oscillation amplitudes" in Table~\ref{table-summary-1}).
Again, the parity symmetry results in the even cat state with these amplitudes,
as shown in Fig.~\ref{fig-single-KPO}(e).

\begin{figure}[t]
	\includegraphics[width=7cm]{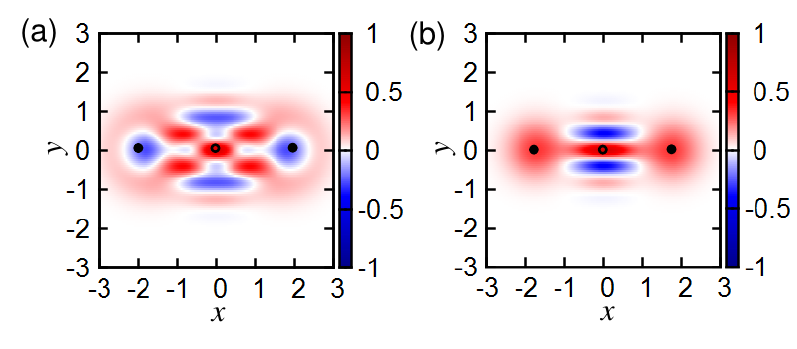}
	\caption{Simulation results for a KPO with a negative detuning.
	(a) and (b) Wigner functions at ${p=3}$ with ${\Delta=-1}$ (a) and ${\Delta=-0.2}$ (b).
	Other parameters are set to the same values as in Fig.~\ref{fig-single-KPO}.
	Filled and open circles represent stable and unstable fixed points, respectively, in the classical model. }
	\label{fig-negative-Delta}
\end{figure}

Here we briefly discuss the case where ${\Delta<0}$.
As mentioned above, this corresponds to the case in Ref.~\citenum{Puri2017a}.
In this case, the Kerr effect decreases the detuning, and hence
does not suppress oscillation amplitudes.
Nevertheless, the Kerr effect stabilizes the oscillation after the amplitudes becomes sufficiently large.
The simulation result for the Wigner function at ${p=3}$ is shown in Fig.~\ref{fig-negative-Delta}(a),
where the parameter setting is the same as in Fig.~\ref{fig-single-KPO} except for ${\Delta=-1}$.
It turns out that we cannot obtain a cat state in this case.
(Instead, we obtain a very intriguing state.
Using such adiabatic processes with negative detunings,
Zhang and Dykman~\cite{Zhang2017b} theoretically proposed a method for 
preparation of intriguing quantum states other than cat states.)

However, if we set $\Delta$ to a smaller value like ${\Delta=-0.2}$,
we can obtain an even cat state~\cite{Puri2017a}, 
as shown in Fig.~\ref{fig-negative-Delta}(b).
This is understood as follows.
In this case,
the vacuum state is the first excited state of the initial Hamiltonian, 
whose ground state is the single-photon state.
Since the quantum adiabatic theorem holds for any energy eigenstate,
the even cat state is obtained via the adiabatic process following the first excited state.
On the other hand, in the case where ${\Delta=-1}$,
the vacuum state and the three-photon state are initially degenerate,
which spoils the adiabatic cat-state generation.

In summary, in the case of negative detunings,
we can control quantum states of a KPO via quantum adiabatic evolution,
but it is necessary to carefully set the detuning to avoid degeneracy.
This is unnecessary in the case of positive detunings.

\section{KPO network}
\label{sec-network}

\subsection{Theory}

Here we first explain how to introduce couplings between KPOs
for solving the Ising problem using a KPO network (QbM).
Next, we explain why the QbM can solve the Ising problem.
Then, the corresponding explanations for an OPO network (CIM) are presented
for comparison.

The Ising problem~\cite{Barahona1982a} requires finding the spin configuration 
that minimizes the following Ising energy:
\begin{align}
E_{\mathrm{Ising}}(\mathbf{s})= 
-\frac{1}{2} \sum_{i=1}^N \sum_{j=1}^N J_{i,j} s_i s_j,
\label{eq-Ising-energy}
\end{align}
where $s_i$ is the $i$-th Ising spin, 
which takes $+1$ (``up") or $-1$ (``down"), 
$N$ is the total number of the Ising spins, 
${\mathbf{s}=(s_1~s_2~\cdots~s_N)}$ is the vector representation of a spin configuration, 
and $J_{i,j}$ is the dimensionless coupling coefficient between the $i$-th and $j$-th spins
(${J_{i,j}=J_{j,i}}$ and ${J_{i,i}=0}$). 
In this paper, for simplicity we do not consider local magnetic fields.
See Refs.~\cite{Sakaguchi2016a,Puri2017a,Goto2018a} for their treatments.

In the case of a KPO network (QbM),
the linear couplings described by the Hamiltonian $\mathcal{H}_c$ in Table~\ref{table-summary-2}
are introduced~\cite{Goto2016a},
where $a^{\dagger}_i$ and $a_i$ are the creation and annihilation operators,
respectively, for the $i$-th KPO 
and $\xi_0$ is a positive constant with the dimension of frequency.

To satisfy the condition that the vacuum state is the ground state of the initial Hamiltonian,
we set $\xi_0$ such that the initial Hamiltonian is positive semidefinite.
This is sufficient for the initial condition,
because the pumping rate $p$ is initially zero and therefore 
the vacuum state is a zero-eigenvalue eigenstate of the initial Hamiltonian.
The condition is satisfied when ${\Delta - \xi_0 \lambda_{\mathrm{max}} \ge 0}$~\cite{Goto2016a},
where $ \lambda_{\mathrm{max}}$ is the maximum eigenvalue of the coupling matrix $J$,
because then the detuning and coupling terms result in a positive semidefinite operator
and the Kerr term is also positive semidefinite.
The physical meaning of the linear couplings is photon exchange between two KPOs.
This is easily implemented by coupling two KPOs directly or 
by using a far off-resonant coupling resonator.
In the latter case, 
we can obtain the linear coupling Hamiltonian 
by adiabatically eliminating the resonator terms.

The corresponding classical model for the KPO network,
which is provided in Table~\ref{table-summary-2},
is derived in the same manner as in the single-KPO case,
The bifurcation point (the threshold for the KPO network) in the classical model
is given by ${p_{\mathrm{th}}=\Delta- \xi_0 \lambda_{\mathrm{max}}}$~\cite{Goto2016a}.
The above condition for a positive semidefinite initial Hamiltonian in the quantum model
corresponds to a nonnegative threshold in the classical model.

When solving the Ising problem, 
the pumping rate $p$ is increased gradually from zero.
Since the initial state is the ground state of the initial Hamiltonian as shown above,
the final state will be the ground state of the final Hamiltonian according to
the quantum adiabatic theorem, provided that the variation of $p$ is sufficiently slow.
If the final value of $p$ is sufficiently large and 
the pumping and Kerr terms are dominant,
the final state will be composed of coherent states, 
as in the single-KPO case.
Thus, the final Hilbert space is approximately spanned by
the $N$-mode coherent states
${| \mathbf{s} \rangle = |{s_1 \alpha_S} \rangle \cdots |{s_N \alpha_S} \rangle}$,
where ${\alpha_S =\sqrt{(p-\Delta)/K}}$ and ${s_j = \pm 1}$
(${j=1,\ldots, N}$).
In this basis, the eigenvalues of the total Hamiltonian 
are given by
\begin{align}
\langle \mathbf{s}|H|\mathbf{s} \rangle
=
\sum_{i=1}^N \left( \frac{K}{2} \alpha_S^4 + \Delta \alpha_S^2 - p \alpha_S^2 \right)
-\xi_0 \alpha_S^2 \sum_{i=1}^N \sum_{j=1}^N J_{i,j} s_i s_j.
\nonumber
\end{align}
Note that the first term is constant and the second term is proportional to the Ising energy
in Eq.~(\ref{eq-Ising-energy}).
That is, the ground state of the KPO network corresponds to the ground state of the Ising model.
Thus, we obtain the solution of the Ising problem from the signs of the final amplitudes.

The Ising problem has two optimal solutions: $\mathbf{S}$ and $-\mathbf{S}$.
Correspondingly, the KPO network has degenerate ground states $|\mathbf{S} \rangle$ and $|{-\mathbf{S}} \rangle$.
Because of simultaneous parity symmetry of the total Hamiltonian~\cite{Goto2016a},
we obtain the \textit{entangled coherent states} (multimode cat states)
$|\mathbf{S} \rangle +|{-\mathbf{S}} \rangle$ via the quantum adiabatic evolution from vacuum states.

In the case of an OPO network (CIM),
couplings are implemented by mutual injection~\cite{Wang2013a}.
(The measurement-feedback technique~\cite{Haribara2015a,Inagaki2016a,McMahon2016a} is not considered in this paper.)
In the classical model,
mutual injection is modeled by adding terms proportional to the coupling coefficients~\cite{Wang2013a},
as shown in Table~\ref{table-summary-2}.
Then the total energy $E$ in Table~\ref{table-summary-2} decreases monotonically.
Since $E$ is of the same form as $H$ for a KPO network,
the threshold for the OPO network is given in a similar manner to that for the KPO network,
as shown in Table~\ref{table-summary-2}.
When the pumping rate $p$ is sufficiently large,
the variables for the steady state are approximated as ${x_j \approx s_j x_{SS}}$ and ${y_j \approx 0}$,
where ${x_{SS} =\sqrt{(p-\kappa)/\kappa_2}}$ and ${s_j = \pm 1}$.
Then, the first term in $E$ is constant and the second term in $E$ is proportional to the Ising energy.
Hence, the signs of the $x$ values for the steady state provide an approximate solution of the Ising problem.

The corresponding quantum model is given by the master equation in Table~\ref{table-summary-2}.
This is reformulated in the standard Lindblad form~\cite{Breuer} as follows:
\begin{align}
\dot{\rho} &= -i[\mathcal{H},\rho] + \mathcal{L}_1 \rho + \mathcal{L}_2 \rho + \mathcal{L}_3 \rho,
\label{eq-Lindblad-1}
\\
\mathcal{H} 
&= i\frac{p}{2} \sum_{i=1}^N \left( a_i^{\dagger 2} - a_i^2 \right),
\label{eq-Lindblad-2}
\\
\mathcal{L}_1 \rho
&= \sum_{i=1}^N \left( \kappa - \xi_0 \sum_{j=1}^N |J_{i,j}| \right) 
\left( 
2 a_i \rho a_i^{\dagger} - a_i^{\dagger} a_i \rho - \rho a_i^{\dagger} a_i
\right),
\label{eq-Lindblad-3}
\\
\mathcal{L}_2 \rho
&= \sum_{i=1}^N \frac{\kappa_2}{2}
\left( 
2 a_i^2 \rho a_i^{\dagger 2} - a_i^{\dagger 2} a_i^2 \rho - \rho a_i^{\dagger 2} a_i^2
\right),
\label{eq-Lindblad-4}
\\
\mathcal{L}_3 \rho
&= \sum_{i=1}^N \sum_{j=1}^{i-1} \xi_0 |J_{i,j}| 
\left( 
2 L_{i,j} \rho L_{i,j}^{\dagger} - L_{i,j}^{\dagger} L_{i,j} \rho - \rho L_{i,j}^{\dagger} L_{i,j}
\right),
\label{eq-Lindblad-5}
\end{align}
where the Lindblad operators for the couplings are defined as 
$\displaystyle {L_{i,j} = a_i - \frac{J_{i,j}}{|J_{i,j}|} a_j}$.
This is implemented by using lossy injection paths~\cite{Takata2015a}, 
where adiabatic elimination of the injection-path terms
leads to the Lindblad operators.
This is a fully quantum-mechanical model for an OPO network (CIM).

Interestingly,
the classical models can also find good solutions of the Ising problem 
with high probability~\cite{Goto2016a}, 
as shown in Fig.~\ref{fig-four-KPO} later.
This is because in the two models, 
the $x$ values become the maximum-eigenvalue eigenvector of the coupling matrix $J$
at the threshold, 
which provides an approximate solution of the Ising problem~\cite{Goto2016a}.
This is easily understood from the fact that the Ising energy in Eq.~(\ref{eq-Ising-energy}) 
is a quadratic form with the coefficient matrix $-J$.
This is an explanation of the minimum-gain principle from a classical point of view.
However, the $x$ values are continuous, unlike Ising spins, and hence 
their dispersion induces errors (trapping at a wrong configuration).
In the quantum model of QbMs,
quantum fluctuations lead to a superposition of many spin configurations at the threshold,
and consequently the system escapes from the trap.
This results in higher performance of the quantum model of QbMs than the classical counterpart~\cite{Goto2016a}
(see Fig.~\ref{fig-four-KPO}).

\subsection{Two coupled oscillators}
\label{sec-two-KPO}

As a simplest example of the Ising problem,
here we consider two coupled oscillators for a two-spin Ising model with ferromagnetic coupling, 
whose ground states are ${s_1=s_2=1}$ and ${s_1=s_2=-1}$.

\begin{figure}[t]
	\includegraphics[width=7cm]{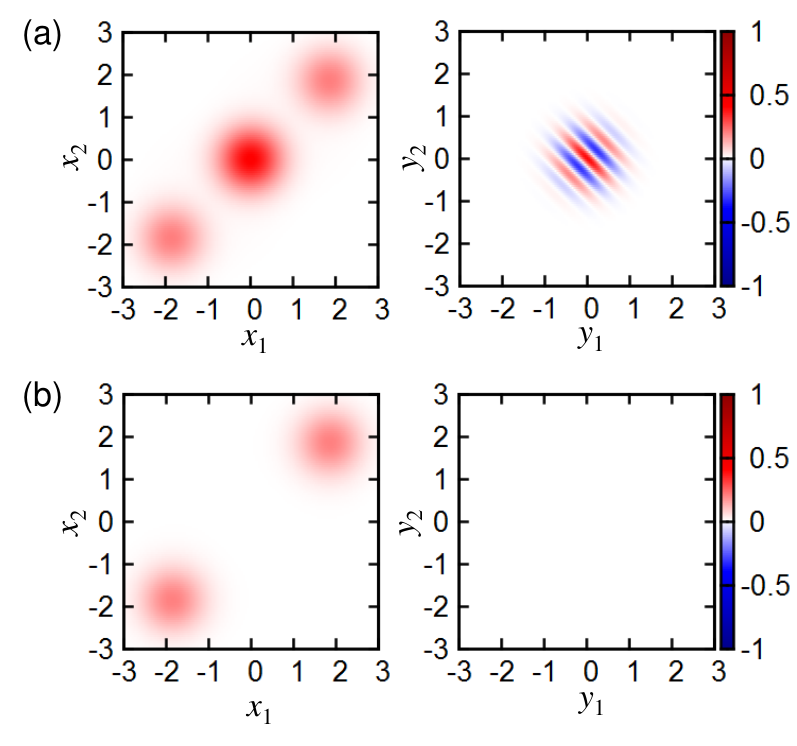}
	\caption{Simulation results for the quantum models of two coupled oscillators.
	(a) Two-mode Wigner function for KPOs at the final time ${t=100}$. 
	Left: ${y_1=y_2=0}$.
	Right: ${x_1=x_2=0}$.
	(b) Corresponding results for OPOs.
	Parameter setting: ${J_{1,2}=J_{2,1}=1}$, $\xi_0=0.5$, and others are set to the same values
	as in Fig.~\ref{fig-single-KPO}.}
	\label{fig-two-KPO}
\end{figure}

Simulation results using the quantum models are shown in Fig.~\ref{fig-two-KPO}.
In this simulation,
we set ${J_{1,2}=J_{2,1}=1}$ and ${\xi_0=0.5}$, and 
the other parameters are set to the same values as in Fig.~\ref{fig-single-KPO}. 
Figures~\ref{fig-two-KPO}(a) and \ref{fig-two-KPO}(b) show the two-mode Wigner functions for the two KPOs and the two OPOs, 
respectively, at the final time.
Here, the two-mode Wigner function is defined as~\cite{Wang2016a} 
\begin{align}
W(\alpha_1, \alpha_2 ) = 
\left( \frac{2}{\pi} \right)^2 
\mbox{Tr} \! \left[ D_1(-\alpha_1) D_2(-\alpha_2) \rho D_1(\alpha_1) D_2(\alpha_2) P_1 P_2 \right],
\nonumber
\end{align}
where $D_i$ and $P_i$ are the displacement and parity operators for the $i$-th oscillator,
respectively.
To display the two-mode Wigner functions with four variables,
we set ${y_1=y_2=0}$ in the left figures 
and ${x_1=x_2=0}$ in the right figures in Fig.~\ref{fig-two-KPO}.

The two peaks around ${x_1=x_2=2}$ and ${x_1=x_2=-2}$ in Figs.~\ref{fig-two-KPO}(a) and \ref{fig-two-KPO}(c)
correspond to the ground states of the Ising model.
That is, both models successfully find the solutions.
However, there are apparent differences around the origin.
There is nothing around the origin in the case of OPOs.
This means that the final state of the two OPOs is a mixed state of two-mode coherent states. 
On the other hand, there is an interference fringe for KPOs.
This shows that the final state of the two KPOs is an entangled coherent state (two-mode cat state),
as expected. 
Such Wigner functions have recently been experimentally observed
using two microwave cavities coupled to a Y-shaped superconducting qubit~\cite{Wang2016a}
by a different technique.

\subsection{Four-spin Ising machines}
\label{sec-QbM}

So far, we have considered four oscillator-network models
(KPO/OPO and quantum/classical).
Hereafter, we use the following abbreviations:
`qQbM' and `cQbM' represent the quantum and classical models, respectively, for QbMs (KPO networks);
`qCIM' and `cCIM' represent the quantum and classical models, respectively, for CIMs (OPO networks).

To evaluate the performances of the four models,
we perform numerical simulation for the four-spin Ising problem with all-to-all connectivity.
We solve 100 instances using the four models, 
where the coupling coefficients $\{ J_{i,j} \}$ for each instance 
are set randomly from among the 21 values $\{-1, -0.9, \ldots, 1\}$.
In this simulation,
${\xi_0=0.25}$ and other parameters are set to the same values 
as in Figs.~\ref{fig-single-KPO} and \ref{fig-two-KPO}.

For the qQbM simulation,
we numerically solve the Schr\"{o}dinger equation in Table~\ref{table-summary-2}.
In the qCIM simulation,
we use a Monte Carlo simulation
called the quantum jump (or trajectory) approach~\cite{Breuer,Carmichael,Plenio1998a}
for the Lindblad-form master equation given by Eqs.~(\ref{eq-Lindblad-1})--(\ref{eq-Lindblad-5}),
instead of solving the master equation directly.
This approach,
which is applicable to any Lindblad-form master equation, 
uses a state vector, instead of a density matrix,
and its implementation is therefore easier and consumes less memory.
We repeat the Monte-Carlo simulation 20 times,
and take their average result.
To simulate the classical models,
we numerically solve the equations of motion in Table~\ref{table-summary-2} $10^3$ times
with initial values set randomly within the interval ${(-0.1, 0.1)}$, and take their average result.

\begin{figure}[b]
	\includegraphics[width=7cm]{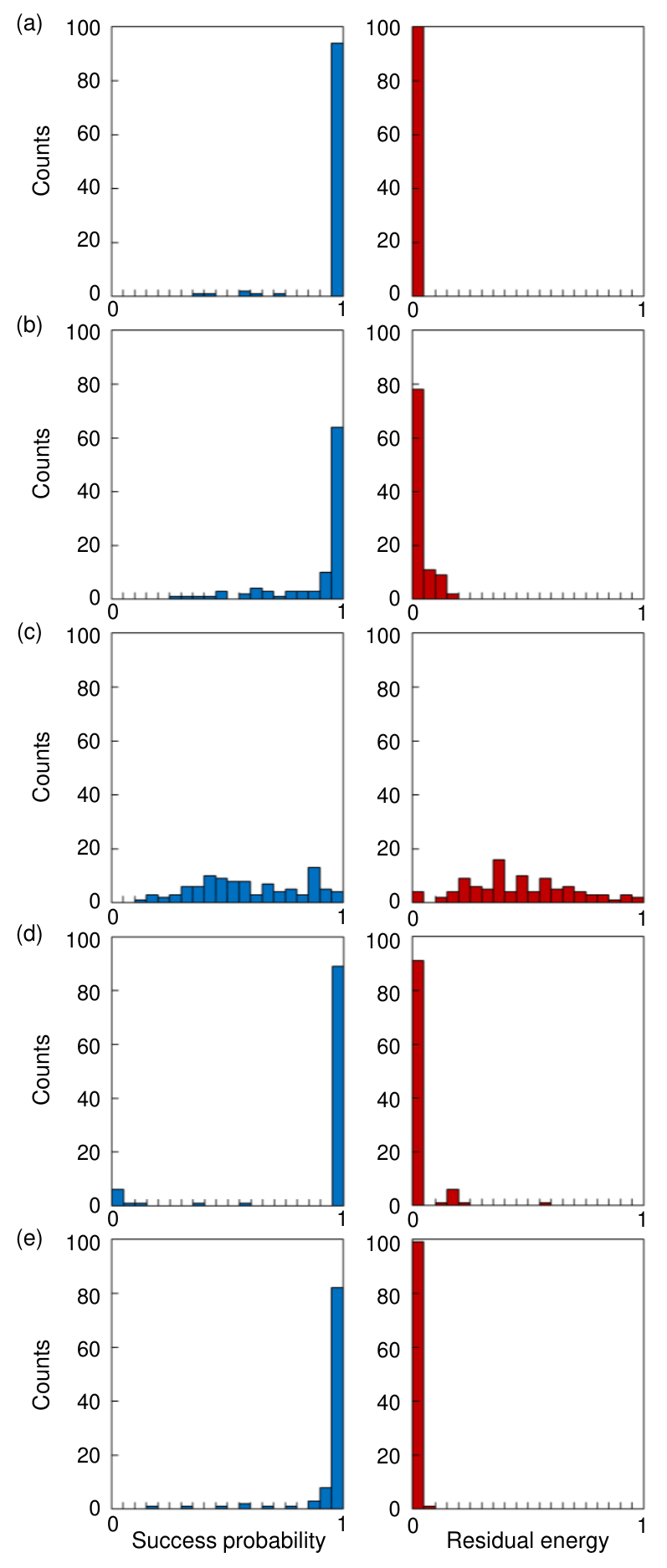}
	\caption{Simulation results for four-spin Ising machines.
	(a) qQbM. (b) cQbM. (c) qCIM. (d) cCIM.
	(e) qCIM without quantum jumps due to one-photon losses in OPOs and injection paths.
	Left: Success probability for obtaining the ground states.
	Right: Residual energy (difference 
	between the average Ising energy obtained in the simulations and the ground-state energy).
	Parameter setting: $\xi_0=0.25$, $\{ J_{i,j} \}$ are set randomly from among $\{-1, -0.9, \ldots, 1\}$, 
	and others are set to the same values 
	as in Figs.~\ref{fig-single-KPO} and \ref{fig-two-KPO}.}
	\label{fig-four-KPO}
\end{figure}

The simulation results are shown by the histograms in Fig.~\ref{fig-four-KPO},
where the probabilities for spin configurations in the quantum models
are calculated using the formula in Ref.~\citenum{Goto2016a}.
Comparing Figs.~\ref{fig-four-KPO}(a)--\ref{fig-four-KPO}(d),
we conclude that qQbM achieves the best performance among the four models.
The high performance of qQbMs as compared with cQbMs is explained by 
quantum superpositions and quantum fluctuations~\cite{Goto2016a}.

The performance of qCIM is remarkably low.
This is due to quantum noises (quantum jumps) from one-photon losses in OPOs and injection paths.
In fact,
its performance becomes much higher under the condition that no such quantum jumps exist, 
as shown in Fig.~\ref{fig-four-KPO}(e).
(This is an unrealistic illustrative condition 
for observing the effects of quantum noise in the present simulation.)

\section{Superconducting-circuit implementations}
\label{sec-SC}

KPOs that can generate cat states have not been experimentally realized so far.
A condition for this realization
is negligibly small dissipation relative to the Kerr coefficient and a parametric pumping rate,
which would be difficult to realize in optical or mechanical systems.
Superconducting circuits with Josephson junctions are natural candidates and the most promising 
for realizing such low-loss KPOs.
Here, we explain how a KPO and Ising machines with KPOs (QbMs) are 
implemented with superconducting circuits~\cite{Puri2017a,Nigg2017a,Zhao2017a}.

\subsection{Implementation of a KPO}

\begin{figure}[t]
	\includegraphics[width=4.5cm]{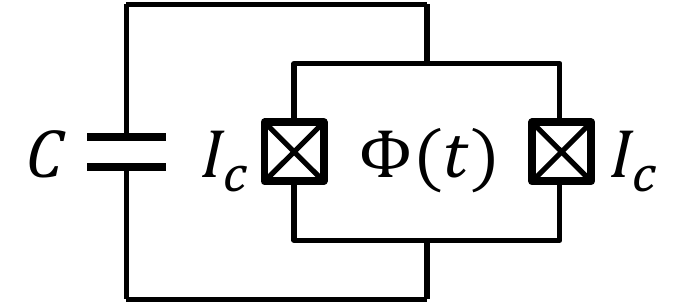}
	\caption{Equivalent circuit for a frequency-tunable transmon qubit.
	$C$: Shunt capacitance for the transmon.
	$I_c$: Critical current for each Josephson junction of the dc SQUID.
	$\Phi (t)$: Magnetic flux through the dc SQUID.}
	\label{fig-transmon}
\end{figure}

As a simplest model for a KPO,
here we consider a frequency-tunable transmon qubit,
an equivalent circuit of which is shown in Fig.~\ref{fig-transmon}.
Transmons~\cite{Koch2007a,Schuster2007a,Majer2007a,Houck2007a,Fink2008a,DiCarlo2009a,DiCarlo2010a,Reed2012a,Barends2013a,Barends2014a,Kelly2015a} 
are widely used superconducting qubits, 
which are composed of a capacitor with large capacitance $C$ and 
a Josephson junction characterized by a critical current $I_c$.
Since a Josephson junction can be regarded as a nonlinear inductor
with ${L_J=\phi_0/(I_c \cos \varphi)}$~\cite{Martinis2004a},
the transmon is a LC resonator with anharmonicity,
where ${\phi_0=\Phi_0/(2\pi)=\hbar /(2e)}$ is the reduced flux quantum ($\Phi_0$ is the flux quantum) 
and $\varphi$ is the phase difference across the junction.
The large capacitance of a transmon leads to a charging energy 
${E_C=e^2/(2C)}$ that is smaller than a Josephson energy ${E_J=I_c \phi_0}$,
making it insensitive to charge noises.
By replacing the Josephson junction with a dc SQUID
(a loop with two identical Josephson junctions),
the critical current can be controlled by the magnetic flux $\Phi$ through the dc SQUID 
as ${\tilde{I}_c=2I_c \cos \! \left( \pi \Phi /\Phi_0 \right)}$~\cite{Tinkham,Makhlin2001a}, 
where $\tilde{I}_c$ denotes the effective critical current for a dc SQUID.
Thus, the resonance frequency of the transmon can be controlled by the flux bias.

The Hamiltonian for a frequency-tunable transmon is given by~\cite{Koch2007a,Martinis2004a,Makhlin2001a}
\begin{align}
H=\frac{Q^2}{2C} - \tilde{E}_J \cos \varphi,
\label{eq-H-transmon}
\end{align}
where $Q$ is the capacitor charge and 
${\tilde{E}_J=\tilde{I}_c \phi_0}$ is the effective Josephson energy for the dc SQUID.
$\varphi$ and $Q$ satisfy the commutation relation $[\varphi, Q]=i(2e)$~\cite{Martinis2004a}.
In fact, using this, 
their Heisenberg equations of motion reproduce the ac and dc Josephson relations~\cite{Martinis2004a,Tinkham} 
as follows:
\begin{align}
\frac{Q}{C} = \phi_0 \dot{\varphi},\quad 
-\dot{Q} = \tilde{I}_c \sin  \varphi.
\end{align}

In the transmon regime (${E_C \ll \tilde{E}_J}$),
the phase difference is confined at the bottom of the potential, 
that is, ${\langle \varphi^2 \rangle \ll 1}$.
Therefore, we approximate $\displaystyle {\cos \varphi \approx 1 -\frac{\varphi^2}{2} + \frac{\varphi^4}{24}}$.
Here we express the dc and ac parts of the flux separately as 
\begin{align}
\Phi (t) = \Phi^{\mathrm{dc}} + \Phi^{\mathrm{ac}}(t),~
\Phi^{\mathrm{ac}}(t) = \delta_p \Phi_0 \cos \omega_p t,
\nonumber
\end{align}
where $\delta_p$ and $\omega_p$ denote the amplitude and frequency, respectively,
of the modulation for parametric pumping.
When $\delta_p \ll 1$,
the Hamiltonian in Eq.~(\ref{eq-H-transmon}) is approximated as
\begin{align}
H &\approx 
\frac{Q^2}{2C} + \frac{\tilde{E}_J^{\mathrm{dc}}}{2} \varphi^2
+ \frac{\tilde{E}_J^{\mathrm{ac}}(t)}{2} \varphi^2
- \frac{\tilde{E}_J^{\mathrm{dc}}}{24} \varphi^4,
\label{eq-H-transmon-app}
\\
\tilde{E}_J^{\mathrm{dc}} 
&\approx 
2 E_J \cos \! \left( \pi \frac{\Phi^{\mathrm{dc}}}{\Phi_0} \right),
\nonumber
\\
\tilde{E}_J^{\mathrm{ac}}(t) 
&\approx 
-2 \pi \delta_p E_J \sin \! \left( \pi \frac{\Phi^{\mathrm{dc}}}{\Phi_0} \right) \cos \omega_p t.
\nonumber
\end{align}

Note that the first and second terms in Eq.~(\ref{eq-H-transmon-app}) 
describe a harmonic oscillator.
Hence $\varphi$ and $Q$ are expressed with creation and annihilation operators as 
\begin{align}
\varphi = \left( \frac{2E_C}{\tilde{E}_J^{\mathrm{dc}}} \right)^{\frac{1}{4}} 
\! \left( a + a^{\dagger} \right),~
Q = i e \left( \frac{\tilde{E}_J^{\mathrm{dc}}}{2E_C} \right)^{\frac{1}{4}} 
\! \left( a^{\dagger} - a \right).
\label{eq-quantization}
\end{align}
Substituting these into Eq.~(\ref{eq-H-transmon-app}) and disregarding c-number terms, 
the Hamiltonian in a frame rotating at $\omega_p/2$ and in the rotating-wave approximation 
is given by the KPO Hamiltonian in Table~\ref{table-summary-1} 
with the following parameters:
\begin{align}
\hbar \Delta &= \sqrt{8E_C \tilde{E}_J^{\mathrm{dc}}} - E_C - \hbar \frac{\omega_p}{2},
\\
\hbar K &= -E_C,
\\
\hbar p &= \pi \delta_p E_J \sin \! \left( \pi \frac{\Phi^{\mathrm{dc}}}{\Phi_0} \right) 
= \frac{\pi \delta_p \tilde{E}_J^{\mathrm{dc}}}{2} \tan \! \left( \pi \frac{\Phi^{\mathrm{dc}}}{\Phi_0} \right),
\end{align}
where $\hbar$ is shown explicitly.
It is notable that the Kerr coefficient is determined only by the charging energy $E_C$~\cite{Koch2007a}
and the one-photon resonance frequency $\omega_{\mathrm{KPO}}$ is approximately given by
the Josephson plasma frequency $\sqrt{8E_C \tilde{E}_J^{\mathrm{dc}}}/\hbar$~\cite{Koch2007a,Tinkham}.

The average photon number for the parametric oscillation with ${\Delta=0}$ is expressed as
\begin{align}
\frac{p}{|K|} 
\approx \frac{\pi \delta_p}{16} \left( \frac{\omega_{\mathrm{KPO}}}{K} \right)^2 
\tan \! \left( \pi \frac{\Phi^{\mathrm{dc}}}{\Phi_0} \right),
\end{align}
where ${\hbar \omega_{\mathrm{KPO}} \approx \sqrt{8E_C \tilde{E}_J^{\mathrm{dc}}}}$ has been used. 
The photon number is upper bounded by the condition 
${\langle \varphi^2 \rangle \ll 1}$.
Using Eq.~(\ref{eq-quantization}),
this condition is rewritten as
\begin{align}
\langle \varphi^2 \rangle
\approx \sqrt{\frac{2E_C}{\tilde{E}_J^{\mathrm{dc}}}} \left( 4 |\alpha|^2 +1 \right) \ll 1
~\Rightarrow~
|\alpha|^2 \ll \! \frac{\omega_{\mathrm{KPO}}}{16|K|},
\end{align}
assuming a coherent state with amplitude $\alpha$
($|\alpha |^2$ is the average photon number).
For typical transmon qubits~\cite{Barends2014a},
$\omega_{\mathrm{KPO}}/(2\pi) \approx 5~\mathrm{GHz}$ and
$|K|/(2\pi) \approx 200~\mathrm{MHz}$, and therefore $\omega_{\mathrm{KPO}}/(16|K|) \approx 1.6$.
This value is too small for parametric oscillations.
This indicates that 
a larger capacitance and a larger critical current are desirable for the implementation of a KPO.

\subsection{Architectures for QbMs}

Here we present two proposed architectures for QbMs~\cite{Nigg2017a,Puri2017a}.

\begin{figure}[b]
	\includegraphics[width=7cm]{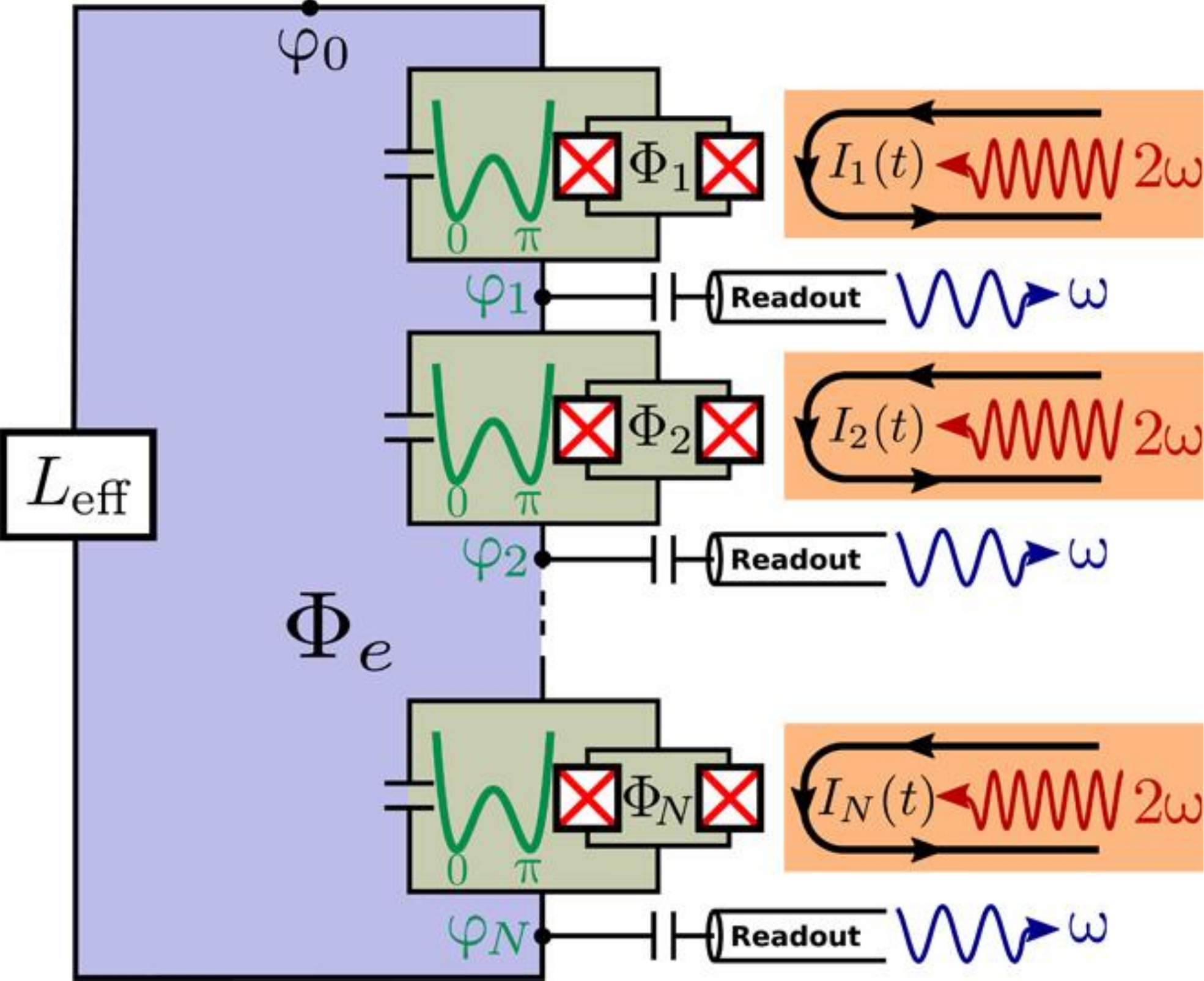}
	\caption{
	KPO-ring architecture for QbM with all-to-all connectivity.
	Adapted from Ref.~\citenum{Nigg2017a} under the CC BY-NC 4.0 license.}
	\label{fig-Nigg}

	\includegraphics[width=7cm]{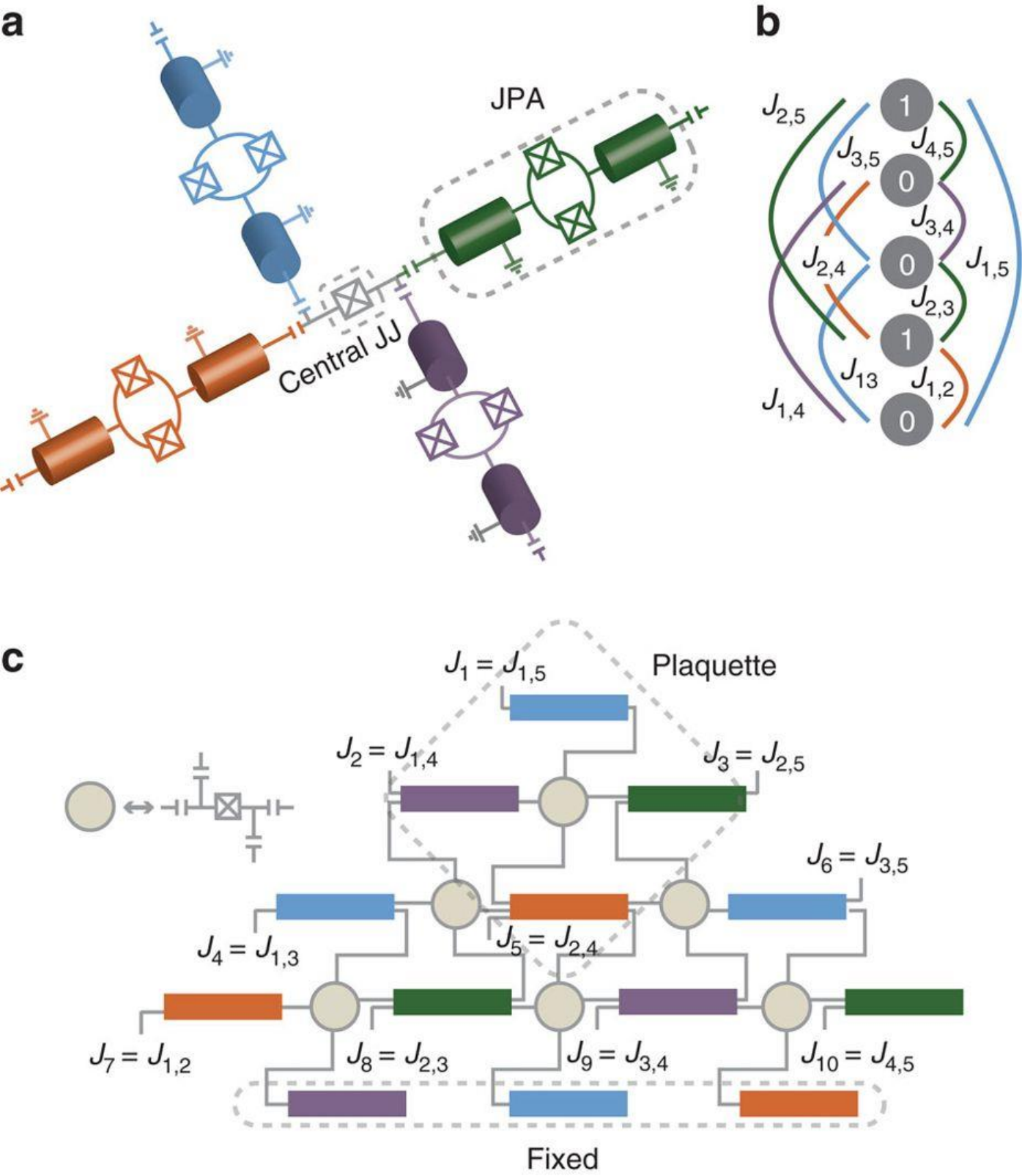}
	\caption{
	QbM architecture based on the LHZ scheme.
	Adapted from Ref.~\citenum{Puri2017a} under the CC BY 4.0 license.}
	\label{fig-Puri}
\end{figure}

Nigg et al.~\cite{Nigg2017a} proposed the KPO-ring architecture shown in Fig.~\ref{fig-Nigg}
for QbMs with all-to-all connectivity.
Antiferromagnetic coupling, which is required for interesting problems,
is realized by setting the flux bias as $\Phi_e = \Phi_0/2$ or 
by shunting the ring with a $\pi$-junction~\cite{Ryazanov2001a,Kontos2002a,Gingrich2016a}.
In this scheme, the coupling coefficients are given by ${J_{i,j} \propto \sqrt{Z_i Z_j}}$,
where $Z_i$ is the mode impedance for the $i$-th KPO. 
Thus, this scheme has only $N$ tunable parameters $\{ Z_1, \ldots , Z_N \}$, 
not all ${N(N-1)/2}$ patterns.
Nevertheless, this can be used for hard problems such as the number partitioning problem,
which is an NP-hard combinatorial optimization problem in which 
$N$ numbers $\{ n_1, \ldots , n_N \}$ 
are partitioned into two groups
such that the sum of one group is equal to that of the other.
The number partitioning problem is equivalent to the Ising problem with ${J_{i,j} \propto n_i n_j}$~\cite{Lucas2014a},
which can be treated in this scheme.
To tune mode impedances to desired values without changing the mode frequencies,
Nigg et al. proposed the use of tunable capacitors.
A scheme to extend the connectivity from ${O(N)}$ to ${O(N \log N)}$
has also been proposed (see the Supplemntary Materials in Ref.~\citenum{Nigg2017a}).

Another QbM architecture with all-to-all connectivity
was proposed by Puri et al.~\cite{Puri2017a} using the LHZ scheme~\cite{Lechner2015a}.
A conventional approach to all-to-all connected quantum annealers
is ``minor embedding."~\cite{Choi2008a,Choi2011a}
The LHZ scheme was proposed as an alternative approach to all-to-all connected quantum annealers 
for realizing a fully two-dimensional layout of qubits
with only nearest-neighbor interactions.
In the LHZ scheme, 
each physical qubit represents the product of an Ising-spin pair, 
instead of an individual Ising spin.
Therefore, this scheme is also called parity adiabatic quantum computing (PAQC)~\cite{Leib2016a}.
Coupling coefficients are introduced as local fields for the physical qubits,
and hence can be controlled more easily than real coupling strengths between qubits.
Corresponding to the coupling coefficients,
this scheme uses ${N(N-1)/2}$ physical qubits.
To reduce the degrees of freedom, 
this scheme requires four-body constraints among four adjacent qubits,
where the eigenvalue of the four-qubit Pauli-$Z$ operator must be one~\cite{Lechner2015a}.
This is an obstacle for quantum annealers based on the LHZ scheme~\cite{Chancellor2017a}.

Remarkably,
the four-body constraint is easily implemented for KPOs
by four-wave mixing in a Josephson junction coupled to four adjacent KPOs,
as shown in Fig.~\ref{fig-Puri}a.
When the four KPOs have different resonance frequencies and 
a four-photon resonance condition, e.g., ${\omega_1+\omega_2 = \omega_3+\omega_4}$ is satisfied,
the four-body interaction yields only terms such as $a_1^\dagger a_2^{\dagger} a_3 a_4$
in the rotating-wave approximation.
These result in the four-body constraint 
in the coherent-state basis~\cite{Puri2017a}.

\section{Universal quantum computation using KPOs}
\label{sec-universal}

A KPO network (QbM) can also be used for gate-based universal quantum computation~\cite{Goto2016b,Puri2017b},
where a qubit is represented by two oscillating states, $|{\pm \alpha_S} \rangle$, of each KPO.
Although ${|\bar{0} \rangle=|{\alpha_S} \rangle}$ and ${|\bar{1} \rangle=|{-\alpha_S} \rangle}$ 
are not orthogonal to each other,
the inner product $\langle {-\alpha_S} |{\alpha_S} \rangle = e^{-2\alpha_S^2}$
is negligible for large $\alpha_S$, and hence 
they can be used for computational-basis states.
Here we briefly explain how to realize a universal gate set for the coherent-state qubits.

A universal gate set is composed of two kinds of single-qubit rotation,
$R_Z (\phi)$ and $R_X(\theta)$,
and a two-qubit gate, $U_{ZZ}(\Theta)$, 
which are defined as follows~\cite{Goto2016b,Nielsen}:
\begin{align}
&
R_Z(\phi) \left( \alpha_0|\bar{0} \rangle + \alpha_1|\bar{1} \rangle \right)
=\alpha_0 e^{-i\phi/2} |\bar{0} \rangle + \alpha_1 e^{i\phi/2} |\bar{1} \rangle,
\nonumber \\
&
R_X(\theta) (\alpha_0|\bar{0} \rangle + \alpha_1|\bar{1} \rangle)
=
\left( \alpha_0 \cos \! \frac{\theta}{2} -i \alpha_1 \sin \! \frac{\theta}{2} \right) |\bar{0} \rangle
\nonumber \\
&
\qquad \qquad \qquad \qquad \quad + 
\left( \alpha_1\cos \! \frac{\theta}{2} -i \alpha_0 \sin \! \frac{\theta}{2} \right)|\bar{1} \rangle,
\label{eq-Rx}
\\
&
U_{ZZ}(\Theta) \left( \alpha_{00} |\bar{0} \rangle |\bar{0} \rangle + \alpha_{01} |\bar{0} \rangle |\bar{1} \rangle
+\alpha_{10} |\bar{1} \rangle |\bar{0} \rangle + \alpha_{11} |\bar{1} \rangle |\bar{1} \rangle \right)
\nonumber \\
&=
e^{-i\Theta/2} \left(
\alpha_{00}  |\bar{0} \rangle |\bar{0} \rangle + \alpha_{11} |\bar{1} \rangle |\bar{1} \rangle \right)
\nonumber \\
&+e^{i\Theta/2}  \left( 
\alpha_{01} |\bar{0} \rangle |\bar{1} \rangle + \alpha_{10} |\bar{1} \rangle |\bar{0} \rangle \right).
\nonumber 
\end{align}

The single-qubit rotation
$R_Z (\phi)$ can be implemented by external driving described by the following Hamiltonian:
\begin{align}
H_{Z}(t) = E_{\mathrm{in}}(t) \left( a+a^{\dagger} \right).
\label{eq-Ein-Hamiltonian}
\end{align}
In the coherent-state basis,
this Hamiltonian is represented as a diagonal matrix
with eigenvalues ${\pm 2E_{\mathrm{in}}(t) \alpha_S}$.
Thus, the two states acquire opposite phases via quantum adiabatic evolution
with a pulse-shaped $E_{\mathrm{in}}(t)$, and hence
$R_Z (\phi)$ is realized.

Similarly,
the two-qubit gate $U_{ZZ}(\Theta)$ can be implemented by 
time-dependent linear coupling:
\begin{align}
H_{ZZ}=g(t) \left( a_1 a_2^{\dagger}+a_1^{\dagger} a_2 \right).
\label{eq-g-Hamiltonian}
\end{align}
Here, 
$|\alpha_S \rangle |\alpha_S \rangle $ and $|{-\alpha_S} \rangle |{-\alpha_S} \rangle $
acquire phases opposite to 
those for $|{-\alpha_S} \rangle |\alpha_S \rangle $ and $|{\alpha_S} \rangle |{-\alpha_S} \rangle $
via quantum adiabatic evolution
with a pulse-shaped $g(t)$, 
and hence $U_{ZZ}(\Theta)$ is realized.

Finally,
$R_X (\theta)$ is implemented by controlling the detuning as follows.
First, Eq.~(\ref{eq-Rx}) is rewritten as
\begin{align}
&
R_X(\theta) 
\left[
\frac{\alpha_0 + \alpha_1}{2} (|\bar{0} \rangle + |\bar{1} \rangle)
+
\frac{\alpha_0 - \alpha_1}{2} (|\bar{0} \rangle - |\bar{1} \rangle)
\right]
\nonumber \\
&
=
\frac{\alpha_0 + \alpha_1}{2} e^{-i\theta/2} (|\bar{0} \rangle + |\bar{1} \rangle)
+
\frac{\alpha_0 - \alpha_1}{2} e^{i\theta/2} (|\bar{0} \rangle - |\bar{1} \rangle).
\label{eq-Rx-2}
\end{align}
Note that 
${|\bar{0} \rangle \pm |\bar{1} \rangle}$
are even and odd cat states.
By increasing the detuning slowly,
even and odd cat states change adiabatically
to vacuum and single-photon states, respectively.
These cat states acquire different phases
depending on the energy gap between the two states during the adiabatic process.
Thus, $R_X (\theta)$ is realized.

\section{Summary and outlook}
\label{sec-conclusion}

We have explained theoretical aspects of Kerr-nonlinear parametric oscillators (KPOs)
and quantum computers with KPOs (quantum bifurcation machines or QbMs),
comparing these with their dissipative counterparts, namely, optical parametric oscillators (OPOs)
and coherent Ising machines (CIMs).
KPOs can generate Schr\"{o}dinger cat states deterministically via quantum adiabatic bifurcations
increasing the pumping rate gradually.
Two coupled KPOs can yield 
entangled coherent states (two-mode cat states).
KPO networks (QbMs) can solve the Ising problem via quantum adiabatic evolution
and also can perform gate-based universal quantum computation.
Superconducting-circuit implementations of a KPO and QbMs have also been presented.
They offer a new application of Josephson parametric oscillators (JPOs).

The first step toward the realization of QbMs is 
an experimental demonstration of the cat-state generation using a KPO.
However, cat states generated inside a KPO are hard to observe directly.
For such observation,
Goto et al.~\cite{Goto2018b}
have recently proposed a method for on-demand generation
of traveling cat states using a KPO.
Since the output field from a KPO can be directly measured,
this method can be used for the experiment.
On-demand generation of traveling cat states has been experimentally demonstrated
using superconducting circuits very recently~\cite{Pfaff2017a}.
The method with a KPO also offers an alternative approach to this challenging task.

Thus, the theoretical proposals of KPOs and QbMs
have opened broad possibilities for theoretical and experimental research 
in the fields of quantum optics, superconducting circuits, and quantum information science.

\section*{Acknowledgments}
This work was partially supported by JST ERATO (Grant No. JPMJER1601).

\end{document}